# A Simple Combinatorial Model of World Economic History

**ROGER KOPPL**
Whitman School of Management
Syracuse University
Syracuse, NY, USA
rogerkoppl@gmail.com

**ABIGAIL DEVEREAUX**
George Mason University
and
New York University
abigail.devereaux@gmail.com

**JIM HERRIOT**
**Herriot Research**
jim@herriot.com

**STUART
KAUFFMAN**
**Institute for Systems Biology**
stukauffman@gmail.com

## ABSTACT

We use a simple combinatorial model of technological change to explain the Industrial Revolution. The Industrial Revolution was a sudden large improvement in technology, which resulted in significant increases in human wealth and life spans. In our model, technological change is combining or modifying earlier goods to produce new goods. The underlying process, which has been the same for at least 200,000 years, was sure to produce a very long period of relatively slow change followed with probability one by a combinatorial explosion and sudden takeoff. Thus, in our model, after many millennia of relative quiescence in wealth and technology, a combinatorial explosion created the sudden takeoff of the Industrial Revolution.

*11 November 2018*

In the classical article "Evolution and Tinkering," François Jacob said natural selection "works like a tinkerer – a tinkerer who does not know exactly what he is going to produce but uses whatever he finds around him whether it be pieces of string, fragments of wood or old cardboards; in short it works like a tinkerer who uses everything at his disposal to produce some kind of workable object." Unlike the engineer, who needs "tools that exactly fit his project," the tinkerer "always manages with odds and ends." Evolution as tinkering has proven a successful metaphor (1, 2). We apply the idea of tinkering where it is not metaphoric: the evolution of technology. We use a simple combinatorial model of technological change to explain the Industrial Revolution.

After the emergence of anatomically modern humans, perhaps about 200,000 years ago (3, 4, 5, 6), income levels changed relatively little until the $19^{th}$ century (C.E.) when rapid technological change produced a spike in per capita incomes first in Europe and North America and then globally. This spike in incomes supported a corresponding spike in global population. (See Figures One and Two). This "hockey stick" of economic growth is the central problem of social science. The Industrial Revolution was the movement from "a traditional world in which incomes of ordinary working people remained low and fairly stable over the centuries" to "a modern world where incomes increase for every new generation" (7). The central question of social science is, then, Why was there a long period of relative stagnation followed by a sudden takeoff producing rapid technological change, sustained growth, and unprecedentedly high incomes? We show that if technological advance is a result of tinkering and recombination, then it may proceed slowly for a long time before a combinatorial explosion generates a rapid increase in the variety of goods and a corresponding increase in wealth.

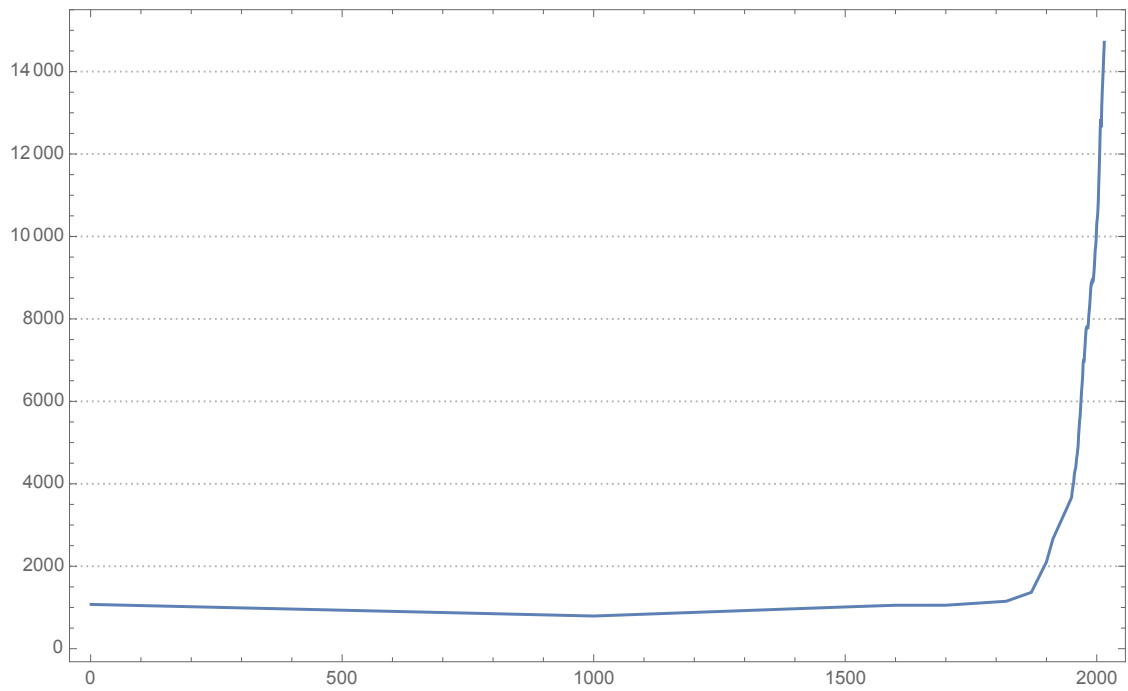

Figure One

GDP per capita over the long run. Calculated using data from (8) and (31,32).

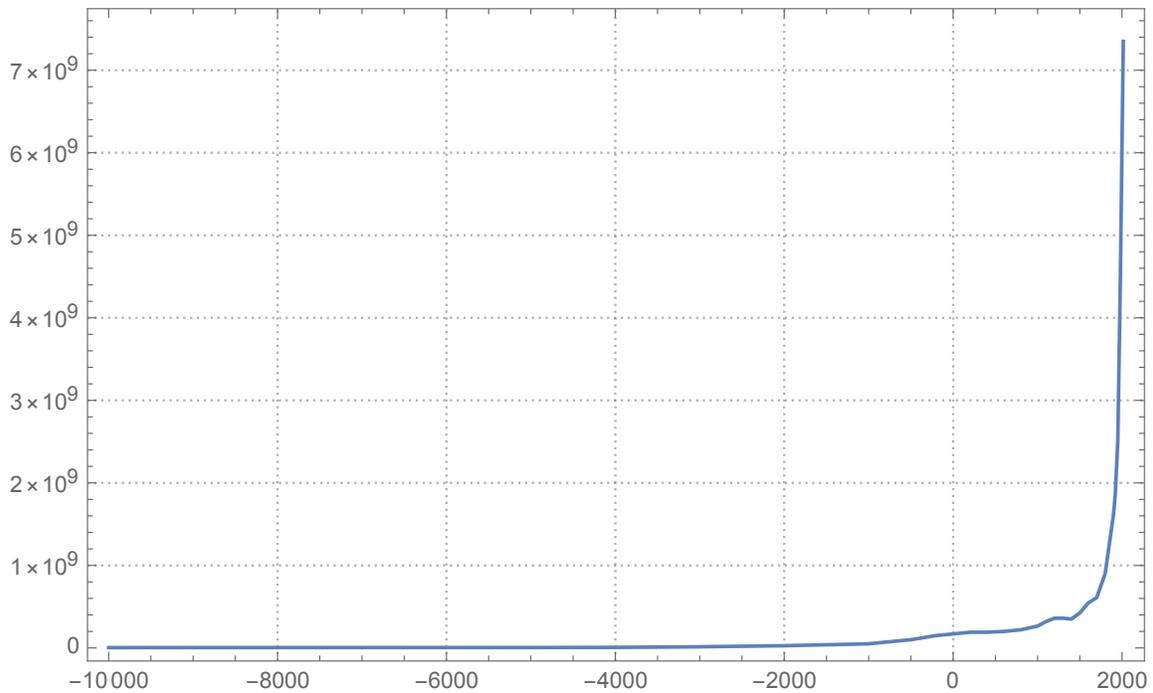

Figure Two

World population from 10,000 BC to the present. Data and estimates derived from (31, 32).

*The industrial revolution*

Prior to the Industrial Revolution global GDP per capita fluctuated between about 450 and 700 1990 Geary-Khamis dollars. Already by 1870 (C.E.) this number had risen to about 870, and thus above historical levels. Global GDP per capita in 2008 was over $7,600, which is about 11 to 17 times larger than historical values. In other words, global per capita GDP today is at least an order of magnitude larger than historical levels. In the richer countries, GDP per capita in 2008 varied between about 20,000 and 30,000 1990 Geary-Khamis dollars (8). Some evidence suggests that our Pleistocene ancestors may have had a standard of life not inferior to historical levels prior to the Industrial Revolution (9). If so, the Industrial Revolution had a far greater effect on incomes than the advent of agriculture and civilization. Some scholars have found

evidence that localized regions such as Northern Italy may have reach relatively high incomes well before the industrial revolution (10-14). Even if all of these recent results were right, however, they would seem to imply little or no change in the basic picture of extended stagnation followed by sudden take off around 1800.

Explanations of the Industrial Revolution include, *inter alia*, exploitation of the worker by the capitalist (15), Calvinism stimulating the emergence of a unique capitalistic form of economic rationality (16, 17), the emergence of trade-friendly institutions initially brought on by England's Glorious Revolution of 1688 (18), the predominance of Christianity in Western culture (19), genetically determined increases in "the average quality in the population" (20), the supposed beneficial eugenic consequences of English primogeniture (9), and a shift in the perceived dignity of commercial activity (21). Each of these explanations appeals to a special cause (or combination of causes) of the Industrial Revolution, and none has emerged as the predominant or consensus view.

We propose a radically different explanation of the Industrial Revolution. Ours is a deflationary explanation. It *deflates* competing views that depend on some special cause or combination of causes to account for the sudden takeoff of the Industrial Revolution. Takeoff might have occurred in another time or place. Its occurrence in late 18$^{th}$ and early 19$^{th}$ century England is largely accidental and independent of any special cultural, institutional, or genetic causes present uniquely in that time and place. In our model, median income is largely unchanging for a long time before a sudden technological takeoff produces increasingly rapid economic growth. We show that this hockey-stick pattern of economic growth can be explained without appeal to any supposed special causes. "Modern growth theory" in economics provides models of the industrial revolution that might also be considered deflationary. We will consider such models after developing our simple model of technological change.

Our model provides a clear mechanism of technological change, which is robust to the institutional context of human exchange. This simple mechanism predates the emergence of anatomically modern man. In our model, the same mechanism of technological change has been operative since the arrival of composite tools about 300,000 years ago, which is probably about 100,000 years prior to the arrival of anatomically modern humans.

*The model*

We model technological progress as increasing "cambiodiversity" (22), that is, as increases in the variety of goods. Increasing cambiodiversity is a characteristic feature of economic growth (23-28). In any period, there is a fixed probability that any one good may be modified to produce a new value-enhancing good, and smaller but still fixed probabilities that 2, 3, or $n$ goods may be combined to produce a new value-enhancing good. Modifications in Paleolithic hand axes (29) and in 17$^{th}$ and 18$^{th}$ century American axes (30) illustrate how an individual good may be modified to produce a new value-enhancing good. Powered heavier than air flight illustrates how two distinct goods – gliders and internal combustion engines – may be combined to produce a new value-enhancing good, the airplane. In our combinatorial model of technological evolution new types of goods may be generated when tinkerers modify an existing good or cobble together two or more existing goods to come up with some new twist or combination that, with all its imperfections and inelegancies, works well enough to be an improvement.

Let $M_t$ denote the number of distinct types of goods in the economy at time $t$. $M_t$ is the degree of cambiodiversity. Our assumption of fixed probabilities of combining $n$ goods to produce a new value-enhancing good leads to the simple combinatorial model given in equation (1) below.

$$M_{t+1} - M_t = P\left(\sum_{i=1}^{M_t} \alpha_i \binom{M_t}{i}\right) \tag{1}$$

where $0 < P\alpha_i < 1$ for $i = 1, 2, \ldots, M_t$ and $\alpha_{i+1} \leq \alpha_i$ for $i = 1, 2, \ldots, M_t - 1$. (In practice, we set $\alpha_i = 0$ for $i > 4$.) $P\alpha_i$ is the probability that if $\binom{M_t}{i} = \frac{M_t!}{i!(M_t-i)!}$ goods are combined they will result in a new good. For simplicity, we take equation (1) to describe the net increase in cambiodiversity rather than separately modeling additions and subtractions to the variety of goods under production.

There is a random element in the emergence of new goods, as reflected in the parameters $P$ and $\alpha_i$. And it may be that many attempts to generate new goods are best viewed as random. But only *value-enhancing* goods will have an enduring place in the econosphere, and it only these value-enhancing goods that we are considering in Equation One.

The fact that only value-enhancing goods will be produced may not be immediately obvious. But if the purpose of production is consumption, then people will not generally have an incentive to engage in consumption-reducing activities. They will not produce a new and innovative good unless it displaces goods of lower value. Production of the new good will consume resources such as human labor that would otherwise have gone into producing other things including, perhaps, leisure. Thus, whenever a value-enhancing good is added to the system, the overall economic output, GDP, goes up. While errors can and will happen, of course, the tendency is always to produce only such innovative new goods as can cover their opportunity costs with at least some surplus.

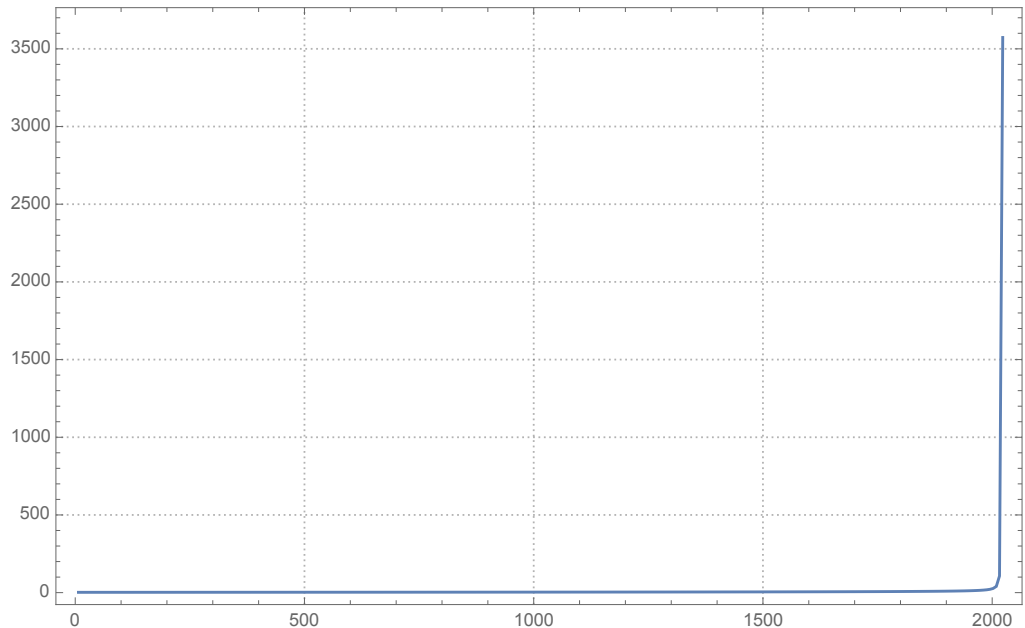

Figure Three

$M_t$ (y-axis), scaled from 0 to 2015 (x-axis).

Our simple model exhibits the characteristic "hockey stick" shape that any explanation of modern economic growth must exhibit. See Figure Three. A long period of largely unchanging values is followed by a sudden takeoff. This behavior emerges from a very simple and unchanging stochastic process. It is not necessary to explain takeoff as the product of some special cause or combination of causes. A low but unchanging value of P will create a long period of stagnation, but sudden takeoff will eventually occur with probability one.

Our model exhibits takeoff in cambiodiversity, which is our measure of technology. Because the rate of increase in cambiodiversity is itself initially increasing, it must produce rapid increases in per capita income, though perhaps with a delay. Similarly, a model with cambiodiversity representing technology exhibits no long-term steady state growth rates of output or capital. Output and capital growth rates are exponential instead of constant. Below we add our model of technological change to a simplified version of a standard "unified model" from the economics literature on "modern growth theory." There is diversity in such models and, more generally, there is an indefinite host of particular mechanisms that might translate technological change into increases in population and per capita income. It seems worthwhile, therefore, to first offer some general considerations to explain why technological takeoff will necessarily bring incomes and population up as well.

With an unchanging population, an increase in cambiodiversity would imply a corresponding increase in per capita income. But if the population is initially at low income levels, then the greater abundance brought about by increased cambiodiversity may be translated into an increase in population size without much changing GDP per capita. And, indeed, before the Industrial Revolution increases in GDP seem to have often, perhaps typically, produced precisely such "Malthusian" income-offsetting increases in population. But technological takeoff produces such a rapid increase in GDP that population cannot rise fast enough to keep up with it. It was thus inevitable that population and average income would both eventually rise precipitously along with cambiodiversity.

Our simple combinatorial model, then, is a sufficient explanation for the Industrial Revolution as defined above. It does not explain the demographic transition occurring after the takeoff in per capita GDP. Prior to the Industrial Revolution, the bulk of the population lived at a relatively low income level, usually dubbed "the subsistence level." In this world, increases in per capita GDP led to increases in population that, in turn, returned per capita GDP to its supposed subsistence level. The rapidly increasing cambiodiversity of the Industrial Revolution caused such rapid increase in GDP that household incomes rose despite population increases. Once incomes rose sufficiently above their subsistence levels, the relationship between income levels and population growth rates changed so that higher incomes now induce falling and not rising population growth rates. This change is the "demographic transition." The rate of world population growth has been declining since about 1970.

*Modern growth theory*

In the "unified models" of "modern growth theory," the system moves endogenously from a "Malthusian" regime of low growth and steady income to a "Post-Malthusian" regime of higher growth and increasing incomes to, finally, a "Modern Growth regime" of continued technological advance in which, however, population growth no longer increases with income, but instead declines (14, 33). In these models, technological change is measured by a scalar whose rate of growth is influenced by the amount of prior knowledge investment. The driver of change may be labeled "education," "R&D," "the number of people engaged in producing ideas," (34), or something else. In this sort of "idea-based theory of [economic] growth," (34), resources are diverted from consumption or other productive activities and invested in knowledge production. These models generally assume that "all knowledge resides in the head of some individual person and the knowledge of a firm, or economy, or any group of people is simply the knowledge of the individuals that comprise it" (35). The mechanism linking such investments to technological change is vague or unspecified.

The following model presents a "unified model" of this type. We recognize that our model is as simplistic as it is simple. Our goal is only to illustrate as simply as possible how our model of technological change can be integrated with existing economic models of growth to generate a "unified model" that conforms with the leading facts of economic history.

In our simple discrete time model $Y_t$ is world GDP in period $t$. $K_t$ is the capital stock in period $t$. It is the value of all goods used to generate, ultimately, final output. $L_t$ is the stock of labor. We assume each living person provides the same quantity of labor, which we normalize to one. Thus, $L_t$ is also the population in period $t$. We measure technology by cambiodiversity, $M_t$. For this simple model, we assume that output is generated by an aggregate production function of the Cobb-Douglas type. Thus,

$$Y_t = M_t K_t^{\beta} L_t^{(1-\beta)}, \qquad (2)$$

where beta is between 0 and 1.

The capital stock, $K_t$, is increased by saving, which we assume to be a fixed fraction, $s$, of output. It is diminished by use as, for example, when machines wear out over time. This "depreciation" occurs at the fixed rate delta. Thus, growth in the capital stock is described by the following equation.

$$K_{t+1} = sY_t + (1-\delta)K_t, \qquad (3)$$

where $s$ and $\delta$ are between 0 and 1.

In the standard economic models of modern growth theory, the population growth rate is derived from the utility maximizing choices of individuals deciding how may children to have and how much to invest in each child. For the sake of simplicity, and to focus on the cambiodiversity, we take population $L_t$ to be exogenous, derived in part from the estimates in (31) and augmented with numbers from the US Census Bureau.

We calibrate the model to reasonably fit growth in total world output from AD 1 to AD 2015, adjusted for inflation and measured in 2011 international dollars (8). We require values for 9 parameters to simulate the model. Following the economics literature, we choose 1/3 for capital's share of output (31, 36) and values $s = 0.25$, delta $= 0.06$.

We assume that $\alpha_i$ is a decreasing function as $i$ increases, and that $\alpha_i = 0$ for $i > 4$. In particular, we assume the decreasing function takes the form

$$\alpha_i = \begin{cases} \frac{1}{(i\theta)^\rho}, & i \leq 4 \\ 0, & i > 4 \end{cases} \tag{4}$$

where $\theta > 0, \rho > 0$. The list of parameters is given in Table One.

| Parameter | Value(s) | Comments |
|---|---|---|
| $Y_0$ | $1.82741 \times 10^{11}$ | Total world GDP at $t = 0$ |
| $M_0$ | 50, 88 | Number of distinct value-adding goods at $t = 0$ |
| $P$ | ~0.0006 | The probability of a successful combination |
| $\theta$ | 6 | $\alpha_i = \frac{1}{(i\,\theta)^\rho}$ is the probability that a combination of $M_t$ goods results in a new good |
| $\rho$ | 2 | $\alpha_i = \frac{1}{(i\,\theta)^\rho}$ is the probability that a combination of $M_t$ goods results in a new good |
| $L_0$ | $1.7 \times 10^8$ | Total world population at $t = 0$ |
| $\beta$ | 1/3 | Capital's share of output |
| $s$ | 0.25 | Fraction of output re-invested into capital formation |
| $\delta$ | 0.06, 0 | Capital depreciation rate |

Table One

Baseline parameter values of the combinatorial growth model, defined by Equations (1), (2) and (3). Entries with comma-delimited values demonstrate more than one good candidate parameter.

Figure Four shows the estimated progression of total world GDP from AD 1 to the present together with simulated values under three different parameterizations. Note that it is simple to extend the simulation backwards from AD 1, by decreasing the initial number of distinct goods $M_0$. We consider it important that the model has validity before AD 1. Importantly, the capital stock $K_t$ should not shrink as output $Y_t$ grows. We chose our parameters to ensure the model is coherent prior to AD 1. The value of $P$ in combination with the parameters $\theta, \rho$ (which determine $\alpha_i$) determine how easy or difficult it is to come up with viable products. A higher $P$, a lower $\theta$, or a lower $\rho$, all else equal, is correlated with a larger $\Delta M_t$ and therefore a larger $\Delta Y_t$.

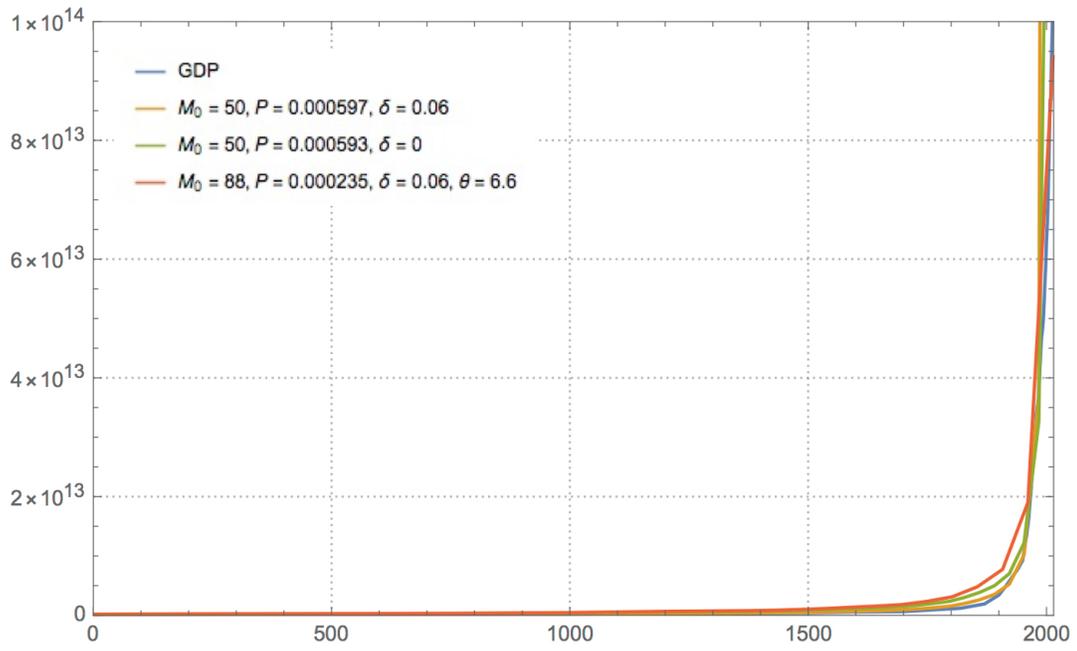
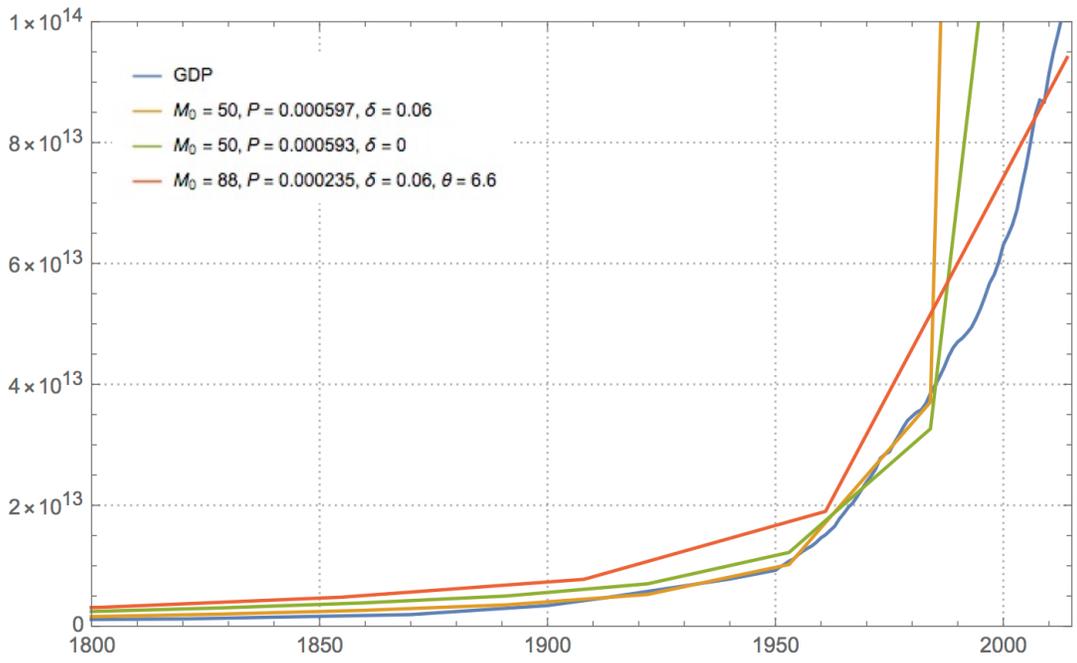

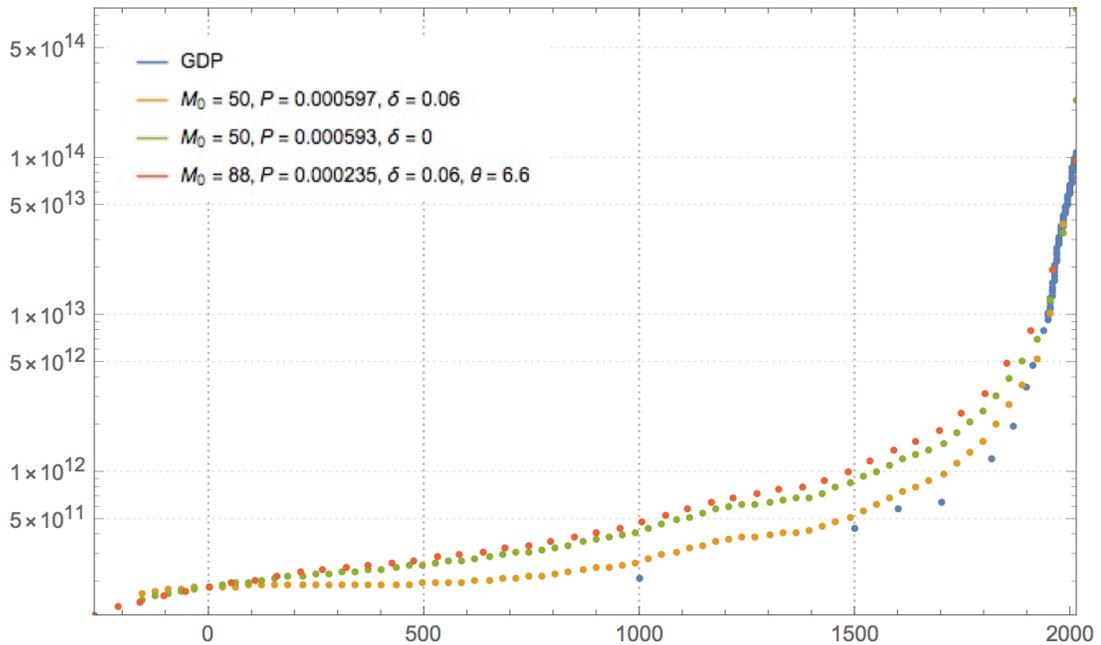

Figure Four

The top graph shows total world GDP (blue) plotted from AD 1 to AD 2015, together with the parameterization $\{M_0 = 50, \delta = 0.06\}$ (yellow), $\{M_0 = 50, \delta = 0\}$ (green), and $\{M_0 = 88, \delta = 0.06\}$ (red). The middle figure is the same graph, zoomed in to AD 1800 to AD 2015, to better visually differentiate between the parameterizations. The bottom figure is the same parameterizations plotted on a log GDP scale, from 350 BC to 2015 AD.

*Discussion*

We have proposed a deflationary theory of the Industrial Revolution. Our theory deflates rivals that rely on some special cause. In our model, the same stochastic process drives technological change from the earliest days to now. There are other explanations that are at least somewhat deflationary, including some of the contributions to modern growth theory discussed above (33, 37). These models, however, neglect the central fact of cambiodiversity. They assume relatively modern institutions of market exchange and are therefore not robust to institutions. They assume self-conscious investments in innovation or research rather than tinkering and chance discovery. And they do not have an explicit and satisfying mechanism of technological change. Finally, we know of only one modern growth theory model that has been shown to generate hockey-stick growth in population, GDP, and income, and this model seems to achieve this result only at the cost of relatively high parameterization (34).

If our basic model of technological change is correct, then the Industrial Revolution would seem to have been the inevitable consequence of the human propensities to tinker, talk and trade. The presence of raw-material transfer distances well above likely maximum territorial radius in the Middle Stone Age suggests that grammatical language and long-distance exchange networks may have emerged 130,000 – 140,000 years ago, (3, 38). The co-evolution of language and trade may have been enabled about 300 kya by the arrival of composite tools, which are "conjunctions of at least three techno-units, involving the assembly of a handle or shaft, a stone insert, and binding materials" (29). We have evidence, therefore, of an autocatalytic process in which increases in cambiodiversity enabled the growth of exchange networks, which, in turn, enabled further increases in cambiodiversity and further growth in exchange networks (24, 26). Mathusian population dynamics prevented increasing cambiodiversity from inducing increases in personal incomes until the combinatorial explosion of technological change finally overwhelmed population growth, thereby inducing sustained increases in per capita GDP.

Our explanation might seem to neglect the important fact of predation, whereby some persons seize (perhaps violently) goods made by others without offering anything in exchange for them. Such "grabbing," as we may call it, discourages technological change. Grabbing in medieval China, where "property was subject to expropriation by Confucian government officials in the name of the emperor," has been used to explain why inventions did not often become innovations in that country (39). A story in Petronius' *Satyricon* vividly illustrates the how grabbing may stifle innovation. An artisan showed Caesar a cup he had made with malleable glass that could not be broken. Once the emperor was satisfied that the artisan had not shared his secret with anyone else, he had the artisan killed on the spot, "for if the secret were known, we should think no more of gold than of mud" (40). In other words, the innovation threatened to drive down the price of the emperor's gold. Grabbing seems to be as ubiquitous in human life as trade. While we do not separately model grabbing, we do not neglect it either. Grabbing reduces $P$, and thus the probability of generating a new good. In a more fine-grained analysis we might attempt to plot the ups and down of $P$ as innovation and grabbing interact and exhibit, perhaps, Lotka-Volterra dynamics. In the end, however, the slow and steady power of even a very low $P$ creates takeoff with probability one. Thus, dropping our simplifying assumption of an unchanging $P$, thereby allowing for differing degrees of grabbing over time, would complicate the analysis without changing the most fundamental contours of our story.

As we have seen, the demographic transition produced a slowdown in population growth. This has been a regime change in population growth. It seems worth inquiring whether incomes and cambiodiversity might not be headed toward similar regime changes. Some evidence suggests that we may be approaching a singularity, perhaps around 2050 (41). We should also recognize, however, the risk of Chicken Littleism. In a classic and important article of 1960, von Foerster et al. estimate that population growth will reach a singularity in 2026 (42). With ironic false precision, they predict "doomsday" on Friday, 13 November 2026. Their humor and irony notwithstanding, they seem to have been sincere in estimating that a population singularity would occur around 2026. As we have seen, however, population growth rates began to fall within about a decade and well before reaching the pitch predicted by their model. It seems, then, both important and difficult to decide whether we will approach a technological singularity, and if so when. Nor is it easy to predict whether the regime change implied by a mathematical singularity would be doomsday or something less dire. In this article, we have adopted the relatively easy task of explaining the past rather than the more daunting task of predicting the future.

*References*

1. Jacob, François. "Evolution and Tinkering," *Science*, 10 June 1977, 196(4295): 1161-1166.
2. Solé, Ricard V., Ramon Ferrer-Cancho, Jose M. Montoya, and Sergi Valverde, "Selection, Tinkering, and Emergence in Complex Networks: Crossing the Land of Tinkering," *Complexity*, 2003, 8(1): 20-33.
3. McBrearty, Sally and Alison S. Brooks, "The revolution that wasn't: a new interpretation of the origin of modern human behavior," *Journal of Human Evolution*, November 2000, 39(5): 453-563.
4. Brown, Francis H., Ian McDougall, and John G. Fleagle, "Correlation of the KHS Tuff of the Kibish Formation to volcanic ash layers at other sites, and the age of early Homo sapiens (Omo I and Omo II)," *Journal of Human Evolution*, October 2012, 63(4): 577-585.
5. Stringer, Chris, "The origin and evolution of *Homo sapiens*," *Philosophical Transactions B*, 5 July 2016, 371(1698): 20150237.